\documentclass[journal]{IEEEtran}
\usepackage[scriptsize]{caption}
\usepackage{graphicx, dblfloatfix}
\usepackage{epstopdf}
\usepackage{amsmath}
\usepackage{amssymb}
\usepackage{graphics}
\usepackage{cite}
\usepackage{amsfonts}
\usepackage{textcomp}
\usepackage{multicol}
\usepackage{multirow}
\usepackage{wrapfig}
\usepackage{color}
\usepackage{fixltx2e}
\usepackage{subfig}

\ifCLASSINFOpdf
\else
\fi
\begin{document}
%
\title{High-Capacity Coherent DCIs using PolMuxed Carrier and LO-Less Receiver}
%
%
%

\author{Rashmi~Kamran,
        and Shalabh~Gupta
\thanks{ Rashmi Kamran and Shalabh Gupta are with the Department
of Electrical Engineering, Indian Institute of Technology Bombay,
Mumbai, 400076 India e-mail: rashmikamran@ee.iitb.ac.in.}
\thanks{Manuscript updated Dec 8, 2018}}

\maketitle

\begin{abstract}
	A PAM\,4 based direct detection system has been standardized for short-distance data center interconnects because of its simple architecture. Performance of the PAM\,4 systems is limited for high dispersion values or demands complicated signal processing for further increase in data rates. A polarization multiplexed carrier based self-homodyne\,(PMC-SH) link with adaptive polarization control is a practical approach with an laser oscillator (LO)-less and carrier phase recovery (CPR)-free coherent receiver that can replace PAM\,4 links for achieving high data rates. We analytically find that PMC-SH scheme results in a significantly better BER for a given transmission rate or can achieve doubling of the data rate for given bandwidth of electronics and laser power (when compared with PAM\,4). Practical implementation of the proposed system with adaptive polarization control is also discussed. Presented theoretical frame work highlights the advantages of such self-homodyne systems over PAM\,4 based systems in terms of SNR requirements and capacity. 
\end{abstract}

\begin{IEEEkeywords}
PAM\,4 links, data center interconnects, self-homodyne system, polarization multiplexed carrier.
\end{IEEEkeywords}
%
\IEEEpeerreviewmaketitle

\section{Introduction}
%
%
%
%
\IEEEPARstart{D}{ata} center applications are dominated by short distance optical links as intra data center links cover 71.6\% of the total applications\cite{cisco3}. Although coherent links with the employment of spectrum efficient techniques like quadrature phase shift keying (QPSK) and quadrature amplitude modulation (QAM) are the first choice for long haul communications, need of power consuming digital signal processing with analog to digital converters make them unsuitable for short distances. A pulse amplitude modulation (PAM\,4) based direct detection system is being used and is widely chosen because of simple hardware and lower power requirements\cite{pamadv} for such data center interconnects (DCI).  As the traffic demands are projected to reach 20.6\,ZB per year by 2021 as per Cisco forecast\cite{cisco3}, the PAM4 adaptability for this future need is a major concern now\cite{pami1,pami2, pami3, edn}. Use of the commercial PAM modules for such high data rates is being discussed in demonstrations for DCI applications\cite{pammicrosoft,pam4demo}. In parallel, implementation of the coherent techniques using remote modulation and self-homodyne (SH) scheme using duplex fiber have also been demonstrated\cite{shduplex,remote}. However, none of the above propositions assures reduction in the signal processing complexity for controlling the power requirements at data centers. For reducing the power consumption and cost, analog domain signal processing has been proposed for coherent receivers \cite{jltn,firstofc,analogcdr}. Presented work investigates an SH scheme to  replace the PAM4 links with the aim of further increasing capacity using analog signal processing based receivers.  
\par
An SH system is a reduced complexity coherent receiver in which the carrier related processing at the receiver is not required as the carrier is sent along with the modulated signal. The SH systems offer many advantages: i) A simple coherent receiver (without local oscillator) and similar receiver electronics is required (without carrier phase noise/carrier phase recovery and correction algorithms that are required in conventional coherent systems). It simplifies electronic signal processing and makes it comparable to the PAM4 links and is feasible for analog domain processing for further power reduction; ii) Phase noise cancellation results in line width tolerance, thus reducing the cost of an expensive laser at transmitter (which is a stringent requirement in conventional coherent systems)\cite{WHYSH4, RPPMC7, RPPMC11}; iii) Opens up a way to increase data rates further by employing spectrally efficient techniques like 16\,QAM and 64\, QAM\cite{16QAM1, SPSH3}. A polarization multiplexed carrier based self-homodyne (PMC-SH) system does not require switches or converters (as required in other SH schemes\cite{RPPMC12,RPPMC13}) that can limit the performance at high speed. Implementation of the PMC-SH systems have been experimentally demonstrated before\cite{RPPMC7, 16QAM1, SPSH3, MPSK, DIS1}. However practical implementation of this technique in DCIs is constrained by the need of continuous polarization control for adequately separating the carrier and the modulated signal at the receiver.
\par The adaptive polarization control for the PMC-SH links can be achieved with an electrically controlled polarization controller (EPC) by minimizing power in one of the polarization at reception\cite{mehula}. Silicon photonics based polarization controllers (PCs) are also being proposed for other applications that can be implemented for PMC-SH systems further\cite{pol2}. Experimental demonstration of a practical PMC-SH system (which uses adaptive polarization control) has been presented in our other work for 64 Gb/s SH-16 QAM system \cite{rashmia1}.
\par In this work, the performance of the PMC-SH links and the PAM4 links respectively is evaluated and compared based on the statistical framework considering all the practical factors like insertion loss and receiver noise. The PAM4 with uniform levels for thermal noise limited receiver is considered. The practical implementation of PMC-SH links is discussed and adaptive PC requirements are also included in the power modeling of the PMC-SH links.

\begin{figure*} [t!]
	\centering
	\includegraphics[scale=0.57]{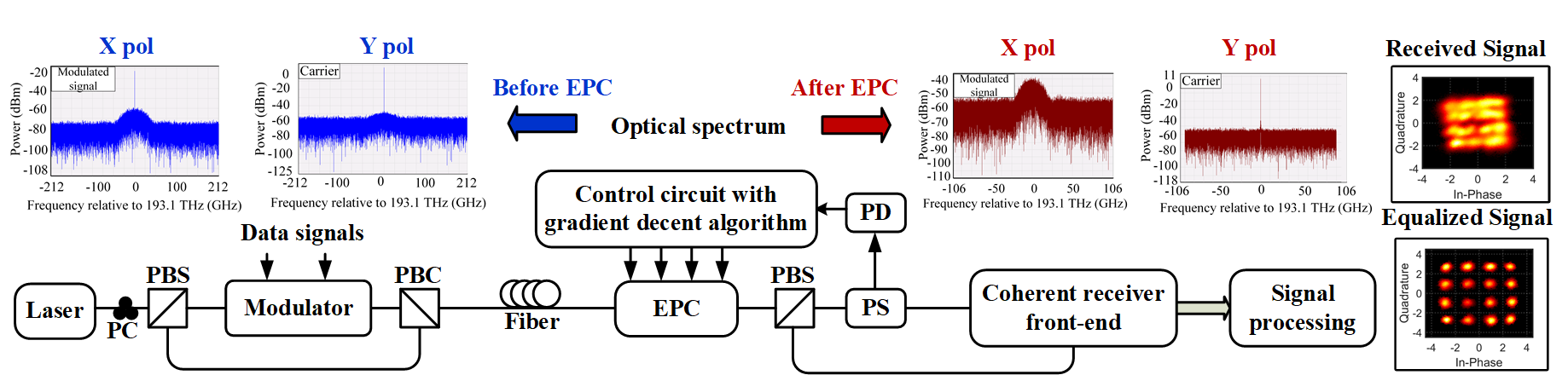}
	\caption{A practical self-homodyne system with polarization multiplexed carrier. PC: polarization controller, PBC/PBS: polarization beam combiner/splitter, EPC: electrically controlled polarization controller, PS: power splitter, and PD: photo detector. Snap shots of the optical spectrums are from the simulation results of 50\,Gbaud SH system with 20\,km fiber. Snap shots of the received signal and equalized signal are from the experimental results of 64\,Gbaud SH-16QAM system with adaptive pol control \cite{rashmia1}.}
	\label{block}
	\vspace{-0.5cm}
\end{figure*} 
\par This paper is organized as follows: Section\,\ref{system} describes system details with applied polarization control method. Section\,\ref{analysis} discusses the detailed statistical framework. Section\,\ref{results} presents the comparison graphs, followed by Section\,\ref{conclusion} which includes conclusion and future work directions.
\section{Proposed system and its practical implementation}
\label{system}
A PMC-SH system block diagram is presented in Fig.\ref{block} with the consideration of its practical implementation with continuous polarization control. Self-homodyning is achieved in this system by using polarization diversity for sending the carrier along with the modulated signal. At the transmitter end, one of the polarization of the laser output is modulated and combined with the other polarization (unmodulated carrier to be used as LO at the receiver). At the receiver end, due to polarization impairments, the modulated signal is mixed with the carrier and the carrier is mixed with the modulated signal. This is the result of random fluctuations in state of polarization by the channel. Adaptive polarization control is required for maintaining the linear state of polarization continuously for the proper separation of two polarization without any cross polarization mixing. Polarization mode dispersion effect is not very significant for the short distance DCIs, so and an EPC with a simple circuitry is able to attain desired state of polarization.
\par For such SH systems, minimization of power in one polarization is able to achieve linear state of polarization as discussed in\cite{mehula}. An electrically controlled PC can be used to maintain minimum power in one polarization by changing its angles based on the control inputs. The control inputs are provided by a control circuit with gradient decent algorithm. Gradient descent algorithm results in control signals towards minimizing the feedback signal. For providing electrical feedback, some fraction ($\approx$10\%) of one polarization power is converted into electrical feedback signal by a photo detector (PD). This control loop successfully separates the carrier and the modulated signal as presented in the snapshots of the simulation results for the optical spectrum's of both polarizations (before and after EPC) in Fig.\ref{block}.  Simulations were performed in VPItransmissionmaker$^{TM}$ for 50\,Gbaud SH-QPSK system with 20\,km fiber. Details of the simulations are available in our previous work\cite{mehula}. 
\par Separated carrier is used as an LO for the coherent receiver front end for the reception of the modulated signal. Receiver front end consists of an optical hybrid and PDs providing electrical I and Q signals. A receiver front end with a monitoring PD can be used to avoid external PD for the feed back that simplifies implementation. Received I and Q signals are applied to signal processor that can be analog processing based equalizer as no complicated processing is required. Equalization is only needed for the dispersion compensation as there is no need for carrier related processing. Practical implementation of this system has been demonstrated by performing experiments for 64\,Gb/s SH-16QAM system with EPC and key results (received signal and equalized signal) are shown in the Fig.\ref{block}. Details of the experimental setup and results are available in our other submitted work \cite{rashmia1}. These experiments validate the practicality of the SH systems in terms of implementation and demands further evaluation in terms of signal to noise ration (SNR) and capacity for short reach links. 
\section{Statistical framework}
\label{analysis}
For evaluating the performance of the PMC-SH links in comparison to the PAM-4 links, expressions for the probability of error P(E) have been derived in terms of laser power and baud rate. The PAM-4 link is compared with the PMC-SH QPSK link (offering same bit rate as PAM-4) and with PMC-SH 16-QAM link (offering double bit rate than PAM-4). Systems have been modeled with the consideration of insertion losses of all the components and for thermal noise limited receivers. Referred basic expressions for all the techniques (PAM-4, QPSK and 16-QAM) have been derived and are detailed in Appendix A.
\begin{figure*} [t!]
	\centering
	\vspace{-0.3cm}
	\includegraphics[width=1.1\columnwidth]{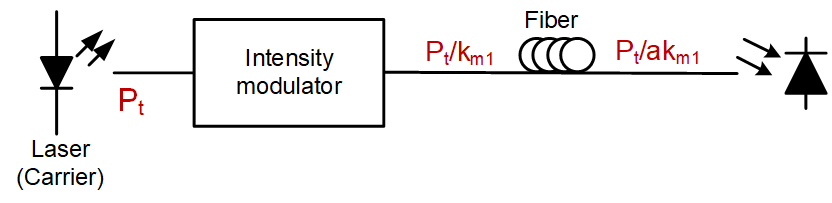}
	\caption{A PAM4 link with received power P$_t$/ak$_{m1}$. P$_t$: laser power, a: attenuation from fiber, and k$_{m1}$: modulator insertion loss. }
	\label{pam4}
	\vspace{-0.5cm}
\end{figure*} 
\begin{figure*} [b!]
	\centering
	\includegraphics[scale=0.45]{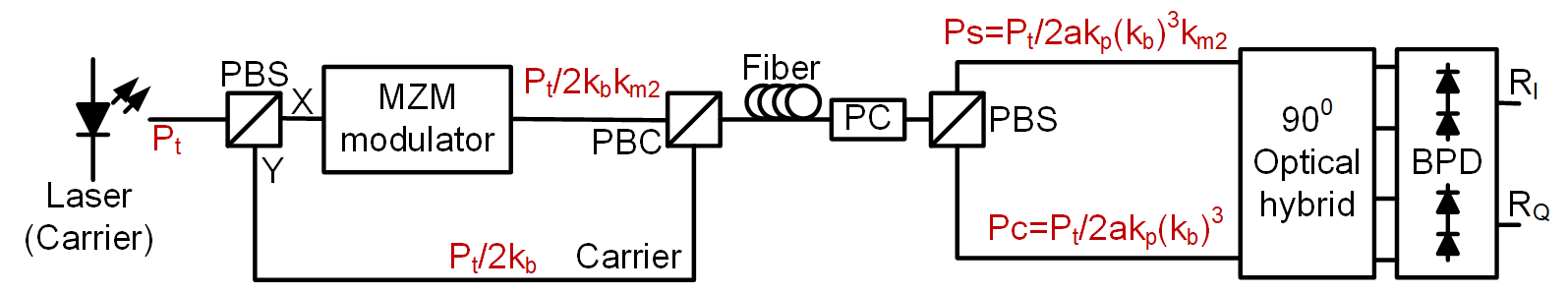}
	\caption{A PMC-SH system block diagram with power modeling at each stage. $k_b$: insertion loss of PBS/PBC, $k_p$: insertion loss of polarization controller, $a$: attenuation due to fiber channel and $k_{m2}$: insertion loss of MZM QPSK modulator.}
	\label{sh}
	\vspace{-0.5cm}
\end{figure*} 

\subsection{PAM4 links}
PAM-4 system has a simple direct detection receiver, as shown in Fig.\,\ref{pam4}, and the received power is affected by the modulator insertion loss and fiber attenuation. General expression for P(E) is detailed in Appendix A in the form of considered amplitude levels $a_1, a_2, a_3, a_4$  for four possible symbols with variances $\sigma_1, \sigma_2, \sigma_3, \sigma_4$ respectively. Uniformly spaced levels are considered for thermal noise limited receiver as $0, \frac{a}{3}, \frac{2a}{3}$ and 1. 
The general expression for BER can be derived as (derivation included in Appendix A):
\begin{equation}
BER_{\text{PAM4}}= \frac{3}{8} \left[\text{erfc}\left(\frac{a}{6\sqrt{2}\sigma_t}\right)\right].
\label{berpam}
\end{equation}
Further $a^2$ in terms of symbol energy ($E_s$) can be expressed as: 
\begin{equation}
\left(0+\frac{1}{9}+\frac{4}{9}+1\right)a^2= 4 E_s,
\label{el}
\end{equation} 
where $E_s$ represents the symbol energy. By keeping $a^2$ in terms of received power and by considering the relation between $a$ and $E_s$ from (\ref{el}) in (\ref{berpam}), effective bit error rate (BER) in the case of thermal noise limited receiver is:
\begin{equation}
BER_{\text{PAM4}}= \frac{3}{8} \left[\text{erfc}\left(0.13\sqrt{\frac{(RP_t)^2}{m(k_{m1})^2n_t^2 \Delta f}}\right)\right].
\label{e1}	
\end{equation}
where $P_t$ is transmitter laser power, $R$ is responsivity of the photodetector, $k_{m1}$ is the insertion loss of intensity modulator, and $m$ is the number of bits per symbol. Thermal noise power is considered as $2n_t^2 \Delta f$, where $n_t$ is thermal noise power spectral density, and $\Delta f$ is the receiver bandwidth.

\subsection{PMC-SH-QPSK links}
Block diagram of a PMC-SH QPSK system along with the power modeling at every stage in terms of laser power is shown in Fig.\,\ref{sh}. The received power is affected by modulator insertion loss and fiber attenuation with additional insertion losses by PBS, PBC, and PC.
\par  At the reception, after separation of both polarizations by a PC, the carrier and the modulated signal is applied to an optical hybrid.  The outputs of the optical hybrid are applied to the BPDs for converting optical signals into I and Q electrical signals. Suppose X polarization is carrying the modulated signal ($\sqrt{Ps}\, e^{j(w_ct+\theta_m)}$) and Y polarization is carrying the carrier signal ($\sqrt{Pc}\, e^{jw_ct}$), where modulated signal power $Ps={P_t}/({2ak_p k_b^3k_{m2}})$ and carrier power $Pc={P_t}/({2ak_pk_b^3})$. Here, $P_t$ is the transmitted laser power, $k_b$ is insertion loss of PBS/PBC, $k_p$ is insertion loss of polarization controller, $a$ is the attenuation due to fiber channel, $k_{m2}$ is the insertion loss of MZM QPSK modulator and $\theta_m$ is phase value according to the modulating signal. The output current from a BPD stage can be expressed as:
\begin{eqnarray*}
	i_1
	&=& R\left[\frac{Ps}{2}+\frac{Pc}{2}+\sqrt{PsPc}\,\cos{\theta_m}\right],\\
	i_2
	&=& R\left[\frac{Ps}{2}+\frac{Pc}{2}-\sqrt{PsPc}\,\cos{\theta_m}\right],\\
	i_1 -i_2 &=& 2R\sqrt{PsPc}\,\cos{\theta_m}.
\end{eqnarray*}
Correspondingly, the thermal noise power is doubled as $4n_t^2 \Delta f$. 
The modulus value of $\cos{\theta_m}$ can be taken as $1/\sqrt{2}$ as all symbols are containing angles multiple of $45^0$. In this case, symbol energy $E_s=2a^2$ as cordinates for QPSK considered are $(a,a), (a,-a), (-a,a)$ and $(-a,-a)$. The general expression for BER can be derived as (derivation included in Appendix A):
\begin{equation}
BER_{\text{QPSK}}=\frac{1}{2}\text{erfc}\left(\frac{a}{\sqrt{2}\sigma}\right).
\label{q1}
\end{equation}
Further by putting the value of $a^2$ in terms of $Ps$ and $Pc$ in (\ref{q1}), resulting expression for BER$_{\text{PMC-SH QPSK}}$ for the case of thermal noise limited receiver is :
\begin{eqnarray}
BER_{\text{PMC-SH\,QPSK}}\hspace{-0.4cm}&=&\hspace{-0.4cm}\frac{1}{2} \text{erfc}\left(\frac{1}{\sqrt{2}}\sqrt{\frac{(RP_t/2)^2}{4m(a k_p k_b^3)^2k_{m2}n_t^2 \Delta f}}\right),\nonumber\\
\hspace{-0.4cm}&=&\hspace{-0.4cm}\frac{1}{2}\text{erfc}\left(0.176\sqrt{\frac{(RP_t)^2}{m(a k_p k_b^3)^2k_{m2}n_t^2 \Delta f}}\right).
\label{e2}
\end{eqnarray}

\subsection{PMC-SH-16\,QAM links}
Links based on 16-QAM technique, offer double data rate as compared to PAM-4 and PMC-SH QPSK links. Block diagram for a PMC-SH 16-QAM link is same as a PMC-SH QPSK system in terms of power modeling. Main differences are the values of insertion loss of the modulators (that is $\approx$ 3\,dB more in the practical system as compared to QPSK modulator) and average energy per symbol. A general expression for P(E) of 16-QAM links with standard uniformly spaced levels is derived as (derivation is included in Appendix A):
\begin{equation}
BER_{\text{16QAM}}= \frac{3}{8}\text{erfc}\left(\frac{a}{3\sqrt{2}\sigma}\right).
\label{qamb}
\end{equation}
Average symbol energy $E_s$ for the standard (uniformly spaced) 16-QAM can be calculated as:  
{	\begin{equation}
	\frac{1}{4} \left[\left(\frac{1}{9}+1\right)+\left(1+1\right)+\left(\frac{1}{9}+\frac{1}{9}\right)+\left(1+\frac{1}{9}\right)\right]a^2=E_s.
	\label{eq2}
	\end{equation}}
Resulting expression for BER$_{\text{PMC-SH 16-QAM}}$ by considering general expressions as in (\ref{qamb}) and (\ref{eq2}) for  the case of thermal noise limited receiver is derived as:
\begin{eqnarray}
 &&\frac{3}{8}\text{erfc}\left(\frac{1}{3\sqrt{2}}\sqrt{\frac{0.9(RP_t/\sqrt{2})^2}{4m(a k_p k_b^3)^2k_{m3}n_t^2 \Delta f}}\right),\nonumber\\
&=&\frac{3}{8}\text{erfc}\left(0.079\sqrt{\frac{(RP_t)^2}{m(a k_p k_b^3)^2k_{m3}n_t^2 \Delta f}}\right).
\label{e3}
\end{eqnarray}
where $k_{m3}$ is the modulator insertion loss for 16-QAM modulation.
\section{Performance evaluation: PAM4 vs PMC-SH systems}
\label{results}
\begin{figure} [h!]
	\centering
	\includegraphics[width=\columnwidth]{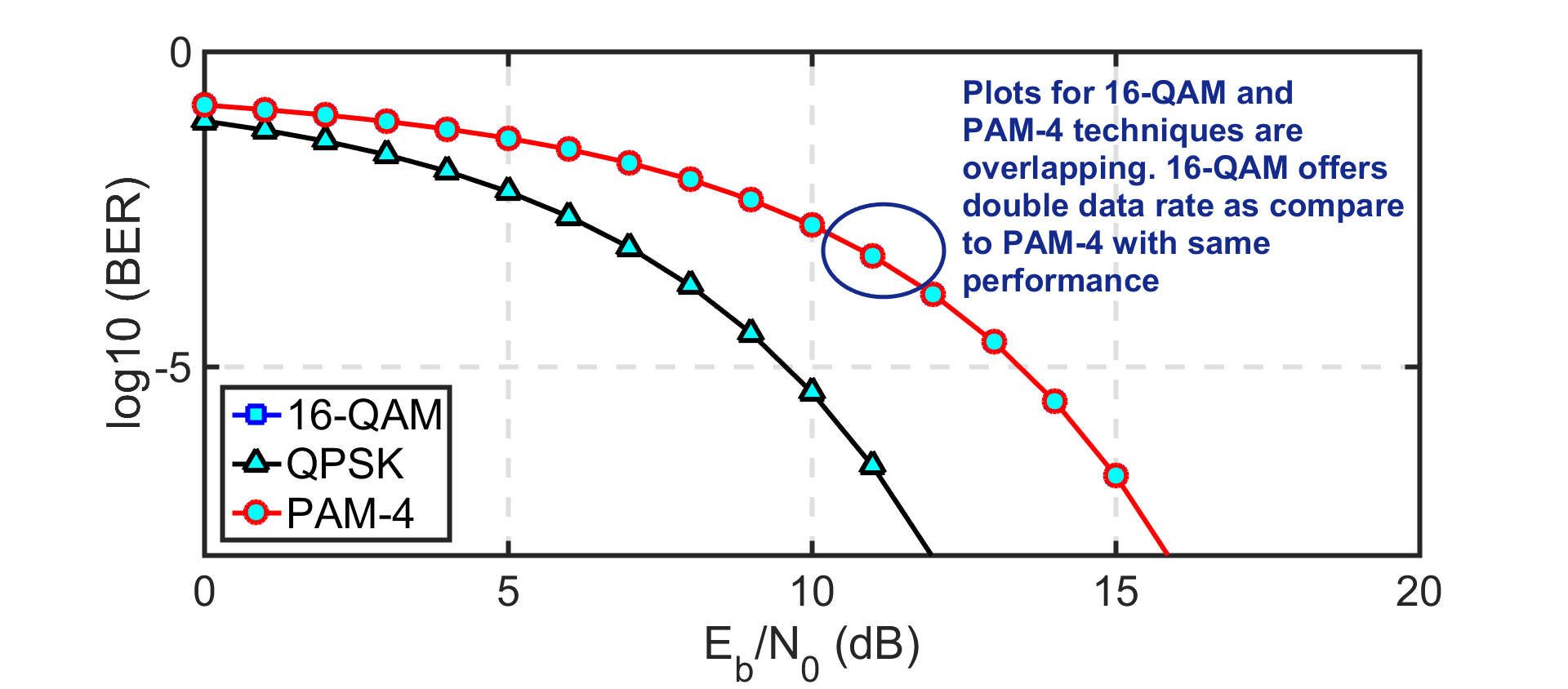}
	\caption{Theoretical ideal BER curve for PAM-4, QPSK and 16-QAM techniques with respect to $E_b/N_0$. This plot is generated by BER analysis application tool of Simulink. }
	\label{ideal}
	\vspace{-0.3cm}
\end{figure}

In the context of evaluating modulation technique performance, graphs are plotted between $E_b/N_0$ and BER to understand the SNR requirement for PAM-4, QPSK, and 16-QAM techniques. This plot is generated by using the Simulink application module for BER analysis. Figure\,\ref{ideal} represents the theoretical performance comparison of modulation schemes without any practical consideration. 
\newline It is showing that QPSK has less SNR requirement and offers the same capacity as PAM-4. Multilevel techniques PAM-4 and 16-QAM have the same SNR requirements for the same performance. Although 16-QAM technique offers double capacity as compared to the PAM-4 technique. This outcome encourages the use of QPSK and 16-QAM techniques for implementation of high-capacity links. Although practical constraints like insertion losses in the system implementation, to be considered further. The purpose of this analysis is to compare the performance of links (PAM-4, PMC-SH QPSK, and PMC-SH 16-QAM) with the consideration of practical constraints.
\begin{figure} [h!]
	\centering
	\includegraphics[width=\columnwidth]{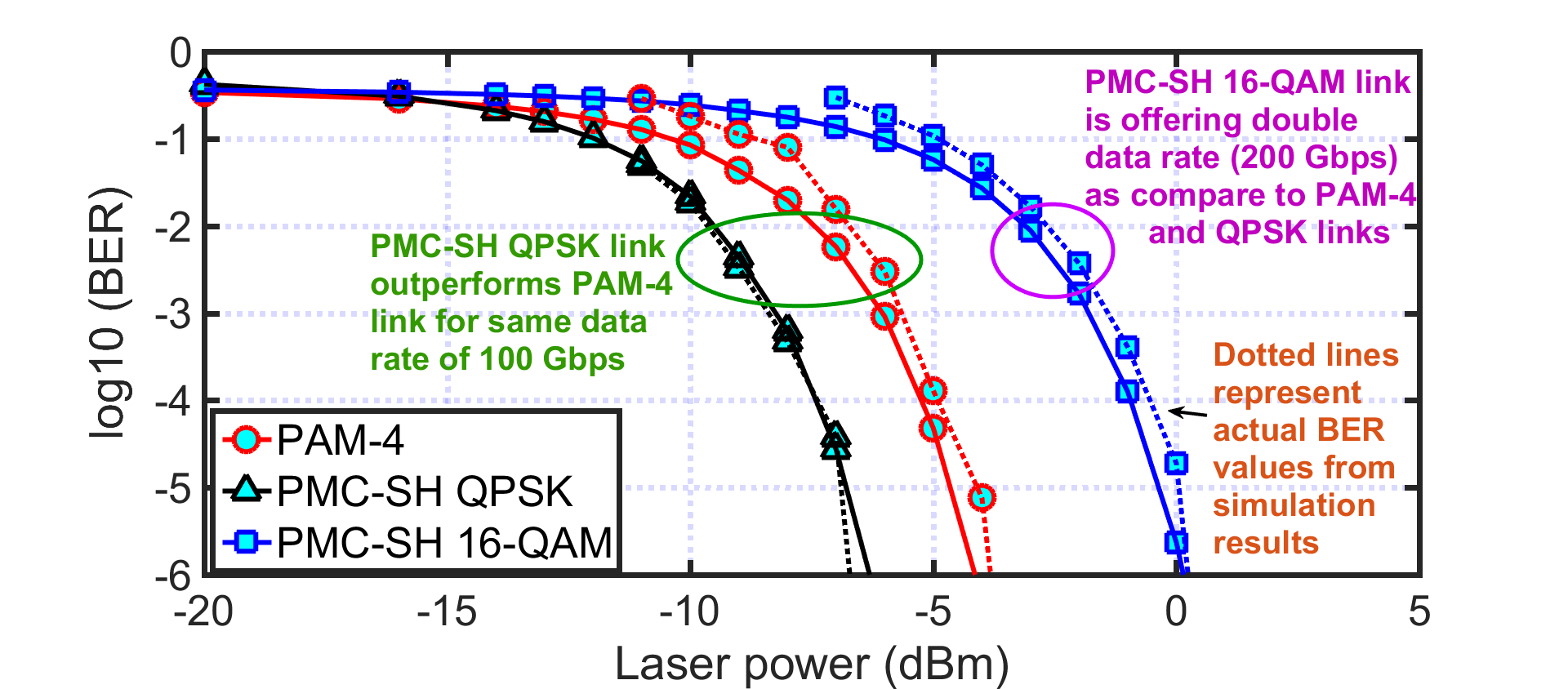}
	\caption{Laser power vs BER for PMC-SH QPSK, PAM-4 and PMC-SH 16-QAM links with thermal noise limited receiver. Parameters values considered are : Responsivity R= 0.85, $\Delta f$= 0.7B where B ia baud rate considered 50\,GSymbols per sec, insertion loss by intensity modulator $k_{m1}$= 7\,dB, noise equivalent power for thermal noise $n_t$ =10\,pW/$\sqrt{\text{Hz}}$,insertion loss by PBC/PBS $k_b$ = 0.5\,dB, insertion loss of modulatior (QPSK) $k_{m2}$= 12\,dB and insertion loss of modulatior (16-QAM) $k_{m3}$= 15\,dB.}	
	\label{pu}
\end{figure}
\par To understand the performance of the systems with practical considerations, expressions presented in (\ref{e1}), (\ref{e2}) and (\ref{e3}) are plotted between the laser power and BER for PAM\,4, PMC-SH QPSK and PMC-SH 16-QAM systems (with the consideration of average power per symbol and system losses) respectively.  Practical values have been considered for all the parameters (values are indicated in the caption of the graph). Multilevel techniques (PAM-4 and 16-QAM) are considered with uniformly spaced levels. By varying laser power, BER is plotted for all three systems in the graphs presented in Fig.\,\ref{pu}. To verify the derived expressions from theoretical analysis, simulations are carried out for all three systems in VPItransmissionmaker. Dotted lines in the graph are showing actual BER values (calculated from simulation results by comparing transmitted and received signals). The overlapping of both graphs (one from the expressions and one from the simulation results) validates the BER expressions derived for practical systems. Corresponding parameter values (as used in theory) and all functional components are considered in simulations. 
\begin{figure} [h!]
	\centering
	\includegraphics[width=\columnwidth]{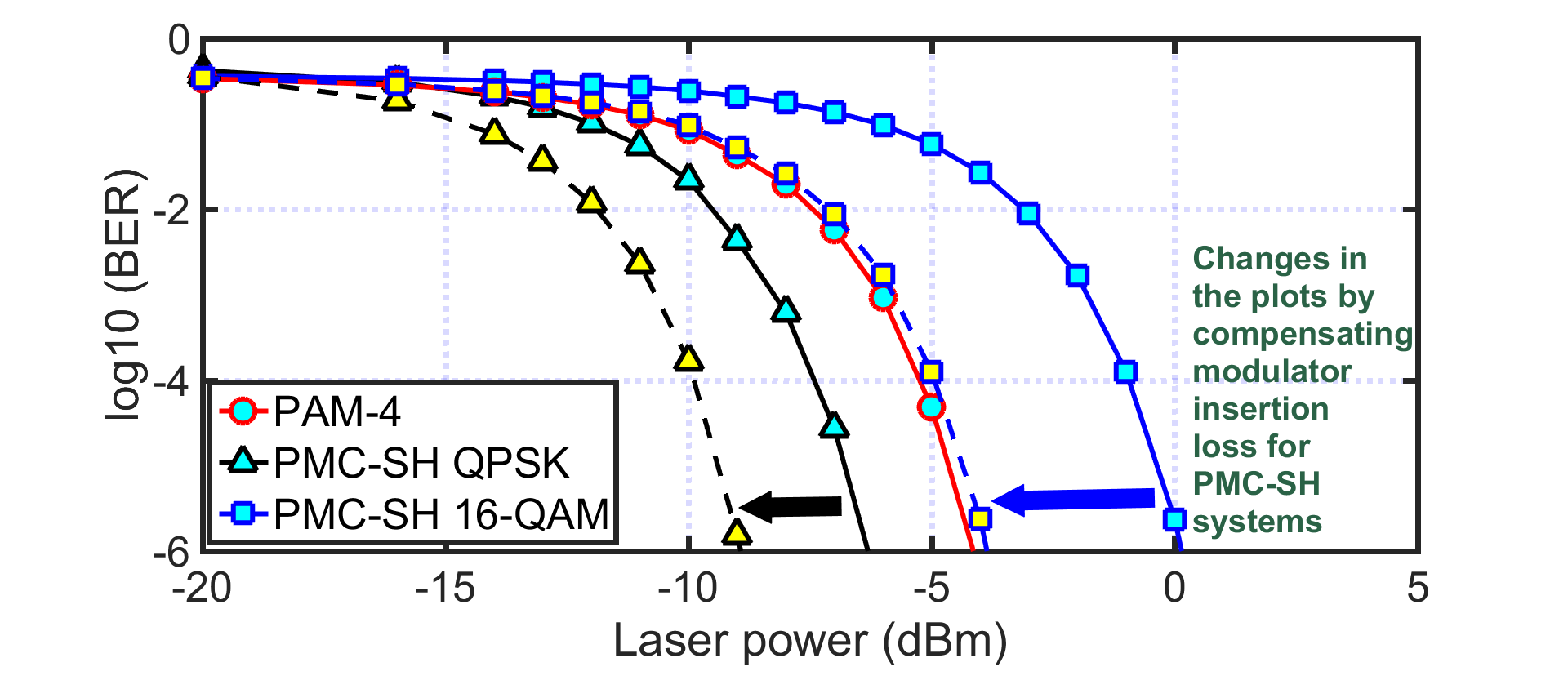}
	\caption{Laser power vs BER for PMC-SH QPSK, PAM-4 and PMC-SH 16-QAM links with thermal noise limited receiver after compensating modulator insertion loss of PMC-SH QPSK and PMC-SH 16-QAM systems (Modulator insertion loss for all systems is taken same as 7\,dB). Parameters values considered are : Responsivity R= 0.85, attenuation from fiber a=5\,dB (by considering 0.5 dB/km for dispersion compensating fiber of length 10\,km), $\Delta f$= 0.7B where B is baud rate considered 50\,GSymbols per sec, insertion loss by intensity modulator $k_{m1}$= 7\,dB, noise equivalent power for thermal noise $n_t$ =10\,pW/$\sqrt{\text{Hz}}$, insertion loss by PBC/PBS $k_b$ = 0.5\,dB, insertion loss by polarization controller and optical hybrid $k_p$ =2\,dB, insertion loss of modulator (QPSK) $k_{m2}$= 7\,dB and insertion loss of modulator (16-QAM) $k_{m3}$= 7\,dB.}	
	\label{compensated}
	\vspace{-0.3cm}
\end{figure}
\par By comparing plot in Fig.\,\ref{ideal} with the plot in Fig.\,\ref{pu}, it can be concluded that insertion loss of the components in the system have increased the SNR requirement for PMC-SH systems specifically for SH-QAM systems. Here modulator insertion loss considered is 7\,dB for intensity modulator, 12\,dB for QPSK modulator and 15\,dB for 16-QAM modulator that is based on discrete component based modulator available. If an optical amplifier compensates modulator insertion loss in PMC-SH systems in the signal path, then this curve can lead to the ideal curve as presented in Fig.\,\ref{ideal}. Figure\,\ref{compensated} is showing BER vs. transmitted power with compensation of modulator insertion loss for PMC-SH systems (modulator insertion loss for all systems are considered equal).

\section{Conclusion}
\label{conclusion}
A PMC-SH-QPSK system outperforms over PAM4 link in every aspect (SNR and required laser power) with offering same bit rate. Performance of a PMC-SH-16\,QAM system is offering double capacity for same bandwidth required for electronics comparable as compared to PAM4 link. This analysis strengthen the employment of PMC-SH links in place of PAM4 links for DCIs.

%

\appendices
\subsubsection{Probability of error for PAM4 links}
\label{p}
\begin{figure} [b!]
	\centering
	\includegraphics[width=\columnwidth]{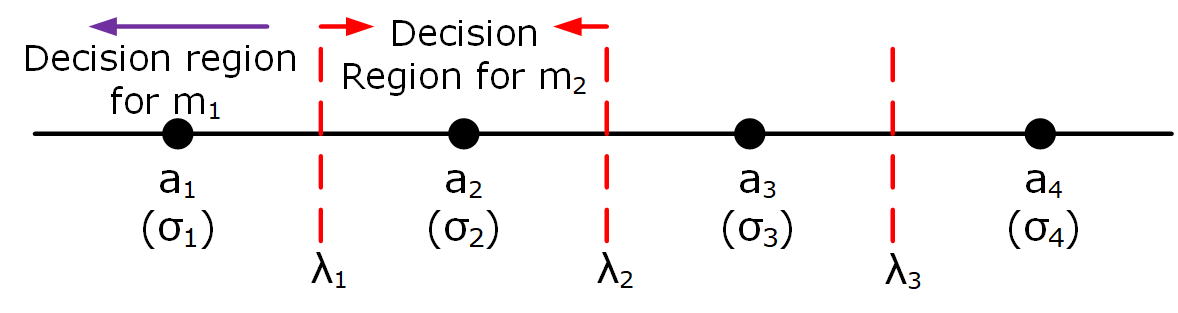}
	\caption{PAM4 constellation diagram. Here amplitude levels $a_1, a_2, a_3, a_4$ are with noise variance $\sigma_1, \sigma_2, \sigma_3, \sigma_4$ and having decision threshold of $\lambda_1, \lambda_2, \lambda_3$. }
	\label{pamc}
\end{figure} 
For considering the effects of non uniform spacing between PAM4 levels and threshold levels to compensate the variance of signal dependent noise, a general constellation as shown in Fig.\,\ref{pamc} is considered. It has four symbols having amplitude levels $a_1, a_2, a_3, a_4$ with noise variance $\sigma_1, \sigma_2, \sigma_3, \sigma_4$ and having decision threshold of $\lambda_1, \lambda_2, \lambda_3$.
Probability of error can be calculated as:
\begin{eqnarray*}
	P(E) &=& p(a_1) P(E|a_1) + p(a_2) P(E|a_2)\\ &+& p(a_3) P(E|a_3) + p(a_4) P(E|a_4),
\end{eqnarray*} 
By assuming equally probable symbols, above expression reduces to: 
\begin{equation*}
P(E) = \frac{1}{4}\left[ P(E|a_1) + P(E|a_2) + P(E|a_3) + P(E|a_4)\right],
\end{equation*} 
Probability of error for symbol $a_1$ can be calculated in terms of probability of correct detection as:
\begin{equation*}
P(E|a_1) = 1- P(C|a_1).
\end{equation*}
Symbol $a_1$ is correctly detected if received symbol lies in the decision area of $a_1$ as indicated in Fig.\ref{pamc}. Probability for error of symbol $a_1$ can be calculated as:
\begin{eqnarray*}
	P(C|a_1) &=& \int_{-\infty}^{\lambda_1} \frac{1}{\sqrt{2\pi \sigma_1^2}} e^{-\frac{(r-a_1)^2}{2\sigma_1^2}} dr \\&=&  1-Q\left(\frac{\lambda_1-a_1}{\sigma_1}\right),\\
	P(E|a_1) &=& Q\left(\frac{\lambda_1-a_1}{\sigma_1}\right).
\end{eqnarray*}
Similarly as per decision area, probability of error for symbol $a_4$ is:
\begin{equation*}
P(E|a_4) = Q\left(\frac{a_4-\lambda_3}{\sigma_4}\right).
\end{equation*}
Probability for correct detection of symbol $a_2$ can be calculated as:
\begin{equation*}
P(C|a_2) = \int_{\lambda_1}^{\lambda_2} \frac{1}{\sqrt{2\pi \sigma_2^2}} e^{-\frac{(r-a_2)^2}{2\sigma_2^2}} dr, 
\end{equation*}
So by solving abaove expression in terms of Q,
\begin{equation*}
P(E|a_2) = Q\left(\frac{a_2-\lambda_1}{\sigma_2}\right) + Q\left(\frac{\lambda_2-a_2}{\sigma_2}\right).
\end{equation*}
Similarly for probability of error for symbol $a_3$ is:
\begin{equation*}
P(E|a_3) =  Q\left(\frac{a_3-\lambda_2}{\sigma_3}\right) + Q\left(\frac{\lambda_3-a_3}{\sigma_3}\right).
\end{equation*}
Total probability of error is:
\begin{eqnarray*}
	P(E) &=& \frac{1}{4}Q\left(\frac{\lambda_1-a_1}{\sigma_1}\right)+ \frac{1}{4}Q\left(\frac{a_2-\lambda_1}{\sigma_2}\right) \\&+& \frac{1}{4}Q\left(\frac{\lambda_2-a_2}{\sigma_2}\right)+ \frac{1}{4}Q\left(\frac{a_3-\lambda_2}{\sigma_3}\right)\\& +& \frac{1}{4}Q\left(\frac{\lambda_3-a_3}{\sigma_3}\right)+\frac{1}{4}Q\left(\frac{a_4-\lambda_3}{\sigma_4}\right).
\end{eqnarray*}
Correspondingly, bit error rate (BER) is (If gray coding is used \cite{haykin}:
\begin{eqnarray}
BER = \frac{1}{4\log_2 M} P(E).
\label{berp}
\end{eqnarray}
\textbf{With uniformaly spaced levels:}\newline
Four amplitude levels $a_1=0, a_2= \frac{1}{3}a, a_3=\frac{2}{3}a, a_4=a$ have been considered for four amplitude levels with same variance for all symbols $\sigma_1=\sigma_2=\sigma_3=\sigma_4=\sigma_t$. Here $\sigma_t$ is the variance representing the thermal noise. Accordingly, threshold levels are $\lambda_1=\frac{1}{6}a, \lambda_2 = \frac{1}{2}a, \lambda_3=\frac{5}{6} a$ as per maximum likelihood criteria. By putting all the values in (\ref{berp}), BER can be calculated as:
\begin{equation}
BER_{\text{PAM4u}}= \frac{3}{4} \left[Q\left(\frac{a}{6\sigma_t}\right)\right]=\frac{3}{8} \left[\text{erfc}\left(\frac{a}{6\sqrt{2}\sigma_t}\right)\right].
\label{berpt}
\end{equation}

\subsubsection{Probability of error for QPSK links}  
\begin{figure} [h!]
	\centering
	\includegraphics[width=0.5\columnwidth]{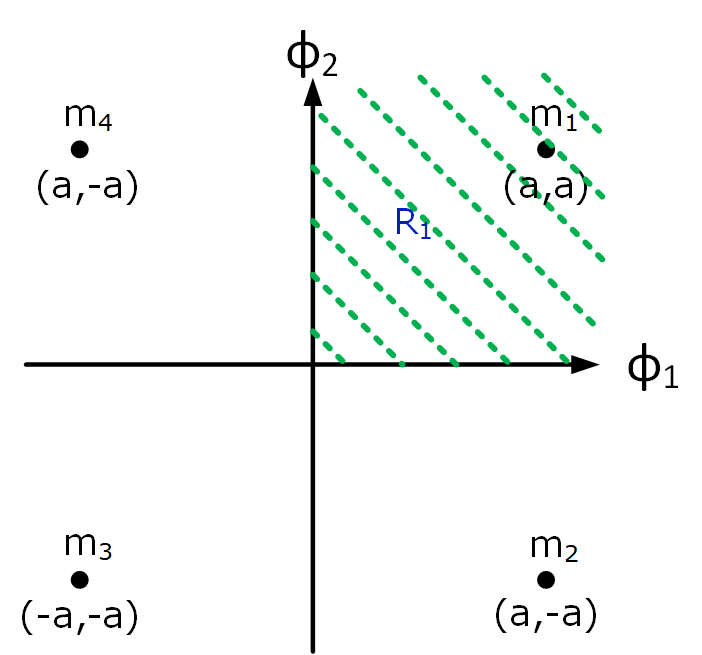}
	\caption {QPSK constellation diagram. }
	\label{qpskc}
\end{figure}

Symbols are considered as per the given constellation diagram in Fig.\,\ref{qpskc} for QPSK links.
Probability of error can be calculated for this techniques also as:
By assuming equally probable symbols, above expression reduces to: 
\begin{equation*}
P(E) = \frac{1}{4}\left[ P(E|a_1) + P(E|a_2) + P(E|a_3) + P(E|a_4)\right],
\end{equation*} 
For QPSK symbols, probability of error is same for all the symbols because of same decision area. So total probability of error can be $P(E)= P(E|a_1)$.
QPSK signal is two dimensional so two received parameters $r_1$ and $r_2$ are considered. These two dimensions are basis functions that are orthonormal. Probability of error for symbol $a_1$ can be calculated in terms of probability of correct detection$P(C|a_1)$ as:
\begin{eqnarray*}
	&=& \int_0^{\infty} \frac{1}{\sqrt{2\pi \sigma^2}} e^{-\frac{(r_1-a)^2}{2\sigma^2}} dr_1  \int_0^{\infty} \frac{1}{\sqrt{2\pi \sigma^2}} e^{-\frac{(r_2-a)^2}{2\sigma^2}} dr_2\\&=&Q\left(-\frac{a}{\sigma}\right) Q\left(-\frac{a}{\sigma}\right),
\end{eqnarray*}
So probability of error is
\begin{equation*}
P(E)=P(E|a_1)= 1-P(C|a_1)= 2Q\left(\frac{a}{\sigma}\right)-Q^2\left(\frac{a}{\sigma}\right).
\end{equation*}
Correspondingly, 
\begin{eqnarray}
BER_{\text{QPSK}}&=&Q\left(\frac{a}{\sigma}\right)- \frac{1}{2}Q^2\left(\frac{a}{\sigma}\right)\nonumber\\&\simeq&\frac{1}{2}\text{erfc}\left(\frac{a}{\sqrt{2}\sigma}\right).
\label{berq}
\end{eqnarray}

\subsubsection{probability of error for 16\,QAM links} 
\label{qam}
QAM is the combination of phase and amplitude modulation. Average power is not same for each symbol so two cases (shot noise limited and thermal noise limited system) are considered for this technique as done for PAM4. This scheme doubles the data rate as compare to PAM4 and QPSK technique based systems.
\begin{figure} [h!]
	\centering
	\includegraphics[width=0.9\columnwidth]{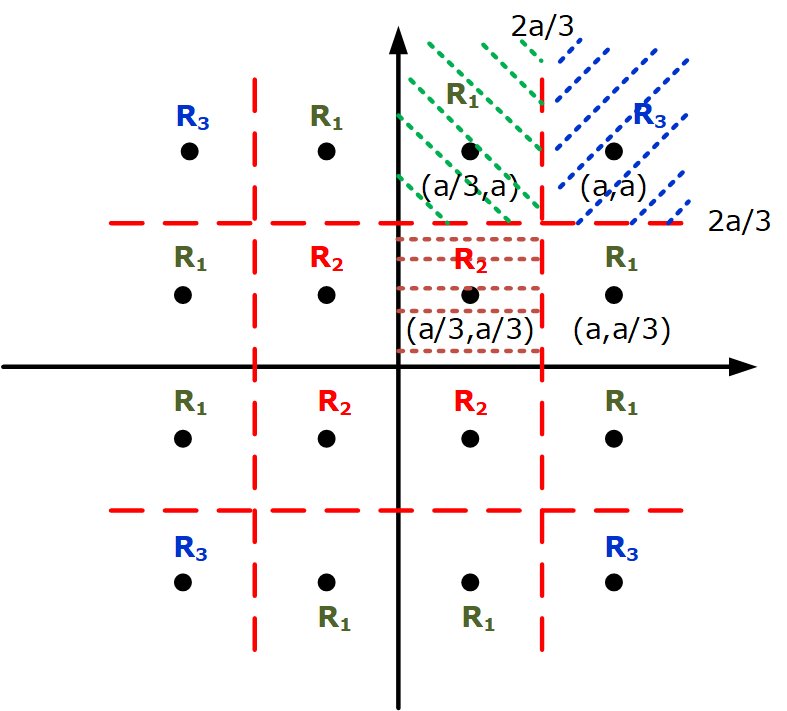}\hspace{1.5cm}
	\caption{16QAM constellation diagram with highlighted decision area. }
	\label{qamc}
\end{figure} 
Signal constellation diagram (in Fig.\,\ref{qamc}) represents the amplitude levels and decision area considered for this case. A signal independent noise is considered with constant noise variance $\sigma$ for all symbols. Symbols having same decision area have similar probability of error. If equally probable symbols are assumed then probability of error is  
\begin{equation*}
P(E)=\frac{1}{16} \left[8P_1 + 4P_2 + 4P_3\right]=\frac{P_1}{2}+ \frac{P_2}{4}+\frac{P_3}{4}, \end{equation*}
where $P_1= P(E|a_1)$, $P_2=P(E|a_2)$ and $P_3=P(E|a_3)$. Firstly $P(C|a_1)$ can be calculated as:
\begin{eqnarray*}
	&=& \int_{0}^{\frac{2a}{3}}\frac{1}{\sqrt{2\pi \sigma^2}} e^{-\frac{(r_1-{\frac{a}{3}})^2}{2\sigma^2}} dr_1  \int_{\frac{2a}{3}}^{\infty} \frac{1}{\sqrt{2\pi \sigma^2}} e^{-\frac{(r_2-a)^2}{2\sigma^2}} dr_2,\\
	&=& \left[1-2Q\left(\frac{a}{3\sigma}\right)\right]\left[1-Q\left(\frac{a}{3\sigma}\right)\right],
\end{eqnarray*}
so,
\begin{equation*}
P(E|a_1)=P_1= 1-P(C|a_1)= 3Q\left(\frac{a}{3\sigma}\right)-2Q^2\left(\frac{a}{3\sigma}\right).
\end{equation*}
Similarly for $P_2$, $P(C|a_2)$ can be calculated as:
\begin{eqnarray*}
	&=& \int_{0}^{\frac{2a}{3}}\frac{1}{\sqrt{2\pi \sigma^2}} e^{-\frac{(r_1-{\frac{a}{3}})^2}{2\sigma^2}} dr_1  \int_{0}^{\frac{2a}{3}}\frac{1}{\sqrt{2\pi \sigma^2}} e^{-\frac{(r_2-{\frac{a}{3}})^2}{2\sigma^2}} dr_2\\&=&\left[1-2Q\left(\frac{a}{3\sigma}\right)\right]^2 ,
\end{eqnarray*}
So,
\begin{equation*}
P(E|a_2)=P_2= 4Q\left(\frac{a}{3\sigma}\right)-4Q^2\left(\frac{a}{3\sigma}\right).
\end{equation*}
For $P_3$, $P(C|a_3)$ can be calculated as following:
\begin{eqnarray*}
	&=& \int_{\frac{2a}{3}}^{\infty}\frac{1}{\sqrt{2\pi \sigma^2}} e^{-\frac{(r_1-{a})^2}{2\sigma^2}} dr_1 \int_{\frac{2a}{3}}^{\infty}\frac{1}{\sqrt{2\pi \sigma^2}} e^{-\frac{(r_2-{a})^2}{2\sigma^2}} dr_2, \\&=&\left[1-Q\left(\frac{a}{3\sigma}\right)\right]^2 ,
\end{eqnarray*}
So,
\begin{equation*}
P(E|a_3)= P_3=2Q\left(\frac{a}{3\sigma}\right)-Q^2\left(\frac{a}{3\sigma}\right).
\end{equation*}
Total probability of error is:
\begin{equation*}
P(E) = 3Q\left(\frac{a}{3\sigma}\right)-\frac{9}{4}Q^2\left(\frac{a}{3\sigma}\right).
\end{equation*}
Correspondingly, BER is (If gray coding is used, BER=SER/$\log_2 M$\cite{haykin}:
\begin{equation}
BER_{\text{16QAM u}}=\frac{3}{4}Q\left(\frac{a}{3\sigma}\right)-\frac{9}{16}Q^2\left(\frac{a}{3\sigma}\right)\simeq \frac{3}{8}\text{erfc}\left(\frac{a}{3\sqrt{2}\sigma}\right).
\label{berqamu}
\end{equation}

\section*{Acknowledgment}
The authors would like to thank Meity for funding the project.




%
\bibliographystyle{IEEEtran}
\bibliography{references18}


%





\end{document}